# Towards Better Summarizing Bug Reports with Crowdsourcing Elicited Attributes

He Jiang, Xiaochen Li, Zhilei Ren, Jifeng Xuan, and Zhi Jin

*Abstract*—Recent years have witnessed the growing demands for resolving numerous bug reports in software maintenance. Aiming to reduce the time testers/developers take in perusing bug reports, the task of bug report summarization has attracted a lot of research efforts in the literature. However, no systematic analysis has been conducted on attribute construction which heavily impacts the performance of supervised algorithms for bug report summarization. In this study, we first conduct a survey to reveal the existing methods for attribute construction in mining software repositories. Then, we propose a new method named Crowd-Attribute to infer new effective attributes from the crowd-generated data in crowdsourcing and develop a new tool named Crowdsourcing Software Engineering Platform to facilitate this method. With Crowd-Attribute, we successfully construct 11 new attributes and propose a new supervised algorithm named Logistic Regression with Crowdsourced Attributes (LRCA). To evaluate the effectiveness of LRCA, we build a series of large scale data sets with 105,177 bug reports. Experiments over both the public data set SDS with 36 manually annotated bug reports and new large-scale data sets demonstrate that LRCA can consistently outperform the state-of-the-art algorithms for bug report summarization.

## I. INTRODUCTION

AS stated in [1], more than 45% of software development effort has been taken on software maintenance for fixing software bugs. Many software projects employ bug repositories, e.g., Bugzilla, to manage numerous bug reports [2]–[5]. For example, Bugzilla has been publicly adopted by 134 organizations and projects, including Mozilla, Eclipse, Gnome, and GCC (www.bugzilla.org/installation-list/). With Bugzilla, over 485,000 and 1,236,000 bug reports have been submitted up to Jan 1, 2017 for Eclipse and Mozilla respectively. Facing numerous bug reports, a lot of software automation tasks have been conducted to facilitate bug fixing [6], including detecting duplicate bug reports [7], [8], triaging bug reports to developers [9]–[11] and locating the root causes of bugs [12]–[14]. During these tasks, people need to well wade through the contents of bug reports. For example, a tester needs to fully understand historical bug reports to avoid submitting duplicate ones [7]. Meanwhile, when fixing a bug, a developer often needs to trace through historical bug reports to locate the root cause of this bug [15] and manually fix the bug with the assistant of bug fixing tools [16]. An earlier work [17] indicates that 200 bug reports from Mozilla contain 275 references to other bug reports, showing the extent to which developers refer to other bugs.

However, it is tedious and time-consuming for testers/developers to wade through the complete contents of bug reports, since a bug report may contain tens even hundreds of sentences [18]. A good way to reduce the time testers/developers take in perusing bug reports is to provide a summary of each bug report [15]. The Debian community even encourages reporters to manually set a summary for each bug report [19], though the considerable human costs may burden this activity. Hence, automatic bug report summarization is an alternative way. Although the title of a bug report is already a good high-level summary [17], [20], the high-level summary is not enough to understand bug reports. For example, testers may hardly detect the duplicate pair of Eclipse bug reports 214301[1] and 214372[2] by merely reading their titles.

In this study, we summarize a bug report by selecting a set of sentences from its description and comments to conclude the main idea of this report. In the literature, a lot of approaches have been proposed to automatically summarize bug reports. The existing approaches can be classified into two categories, namely supervised [15], [21] and unsupervised ones [17], [18]. Given a bug report, both categories evaluate all the sentences in this bug report and select some of them to form its summary.

In a supervised approach, a set of attributes[3] characterizing the sentences in bug reports are constructed and evaluated to train a statistical model over a training set (annotated bug reports in this task). Given a new bug report, the attributes of its sentences are calculated and fed into the trained model to produce its summary. In the seminal work [15], Rastkar et al. first issue the task of bug report summarization and propose a supervised approach named Bug Report Classifier (BRC). They directly transfer 24 attributes from generic conversation-based summarization to bug report summarization and train the model of logistic regression to predict the summary sentences, i.e., the sentences in the bug report summary [22].

In contrast, an unsupervised approach usually assigns a

H. Jiang is with School of Software, Dalian University of Technology, Dalian, China, and Key Laboratory for Ubiquitous Network and Service Software of Liaoning Province. E-mail: hejiang@ieee.org

X. Li, and Z. Ren are with School of Software, Dalian University of Technology, Dalian, China. E-mail: li1989@mail.dlut.edu.cn, zren@dlut.edu.cn

J. Xuan is with School of Computer Science, Wuhan University, Wuhan, China. Email: jxuan@whu.edu.cn

Z. Jin is with Key Laboratory of High Confidence Software Technologies (Peking University), Ministry of Education. Email: zhijin@pku.edu.cn

---

[1]Bug 214301 - Could not load tasklist hyperlink detector extension. https://bugs.eclipse.org/bugs/show_bug.cgi?id=214301

[2]Bug 214372 - Exception during editing xsd files. https://bugs.eclipse.org/bugs/show_bug.cgi?id=214372

[3]In the societies of data mining and machine learning, attributes are also known as features.

measure value to each sentence in a bug report and selects the top ranked sentences to form the summary. Mani et al. [18] apply four unsupervised algorithms to this task, namely Centroid [23], Maximum Marginal Relevance (MMR) [24], Grasshopper [25], and Diverse Rank [26]. Lotufo et al. [17] propose an unsupervised approach by analyzing how developers scan a bug report.

In this study, we focus on the supervised approaches for bug report summarization and investigate new methods to construct effective attributes to facilitate this task. In Mining Software Repositories (MSR), attributes, namely the data representation ways with discriminative information from data, heavily impact the effectiveness of supervised approaches, since different attributes can "entangle and hide more or less the different explanatory factors of variation behind the data" [27]. However, to the best of our knowledge, no analysis has been systematically performed to investigate the methods behind constructing new attributes for bug report summarization (also for other tasks in MSR). To achieve more insights into this topic, several issues are to be investigated. In addition to knowledge transfer employed in [15], what are the other methods adopted in MSR for constructing attributes? Can we have better methods to construct effective attributes? What is the performance of supervised approaches with such new attributes against existing ones?

In this study, *we first conduct a survey on the authors of 40 papers in MSR to reveal the existing methods for constructing attributes*, including knowledge transfer, mining data, and heuristic or experience (see Section III for more details). Meanwhile, this survey indicates that mostly 1-3 persons are involved in the process of constructing attributes, which may bring some potential drawbacks. On the one hand, knowledge transfer is inapplicable unless researchers identify similar domains with effective attributes. On the other hand, it is hard for the other two methods to achieve enough effective attributes when a limited number of researchers are involved.

Second, *we propose a new method named Crowd-Attribute (CA) to systematically infer attributes for bug report summarization from the crowd-generated data in crowdsourcing*. Although crowdsourcing has been exploited in many Software Engineering (SE) tasks [28], [29], as to the best of our knowledge, CA is the first attempt towards employing crowdsourcing to generate data for attribute inference. CA holds a series of characteristics. On the one hand, CA involves a group of volunteers to inspire researchers so it is hopeful to achieve more effective attributes. On the other hand, new attributes by this method are able to achieve promising performance in summarizing bug reports. More specifically, CA chooses a part of bug reports and employs a group of volunteers to manually extract summary sentences. During the process, the volunteers are requested to report the reasons in making their decisions. Inspired by these reasons, we construct new attributes under the guidance of a set of Heuristic Construction Rules. In this study, we develop a tool named Crowdsourcing Software Engineering Platform (CSEP) to facilitate this process.

Third, *we successfully construct 11 new attributes by applying the new method CA and propose a new approach named Logistic Regression with Crowdsourced Attributes (LRCA) to*

*achieve more accurate summaries of bug reports*. In LRCA, the 11 attributes are evaluated to train a statistical model, namely logistic regression, over the training set. For a new bug report, the attributes of each sentence are calculated and fed into the trained model to predict its summary.

Finally, *we conduct extensive experiments to evaluate the effectiveness of LRCA with a series of data sets*. In addition to the only publicly available data set SDS (the bug report Summarization Data Set) [15] with 36 annotated bug reports, we also build a series of large data sets named Bug Report Summarization Benchmarks (BRSBs) with 105,177 bug reports from four well-known open source projects. Over the manually annotated data set SDS, LRCA improves the supervised approach BRC by 1.33%, 10.11%, 8.94%, and 5.89% in terms of *Precision*, *Recall*, *F-score*, and *Pyramid* respectively. Meanwhile, LRCA consistently improves the unsupervised approaches by 0.93%-18.77% over these metrics. For BRSBs, LRCA improves BRC by 22.44% - 34.72%. In contrast, LRCA can also improve the unsupervised approaches by 1.88% - 29.4%.

This paper is structured as follows. In Section II, we present the background and the motivation of this study. In Section III, we investigate the existing methods for constructing attributes in the literature by a survey. In Section IV, we illustrate the roadmap of CA and the new tool CSEP. Then, we apply CA to bug report summarization in Section V. Experimental results are illustrated in Section VI. We conduct a discussion on CA and experimental results in Section VII. The threats to validity are presented in Section VIII. We review the related work in Section IX. Finally, Section X concludes this paper.

## II. BACKGROUND AND MOTIVATION

In this section, we first present the background of bug report summarization, then clarify related conceptions of attributes in supervised approaches to justify the motivation of this study.

### A. Bug Report and Bug Report Summarization

Fig. 1 exhibits a bug report of Eclipse, namely Bug 214066[4]. As shown in Fig. 1, a developer named Steffen Pingel submits a bug report entitled "improve switching between task list presentations". Steffen Pingel presents the detailed description of this bug that "It is not possible ... mode)". In addition, Steffen Pingel also specifies some related items of this bug, e.g., *Product*, *Component*, etc. After Steffen Pingel initializes Bug 214066, five participators comment on this bug.

Given a bug report as Bug 214066, the goal of bug report summarization is to extract a part of sentences from its description and comments to form the summary which can present the essence of this bug report. Up to now, both a supervised approach [15] and several unsupervised approaches [17], [18] have been proposed to resolve this task. To date, SDS is the only widely-used and publicly available data set for this task [15]. In [15], Rastkar et al. collect 36 bug reports with more than 2,000 sentences from four open source software projects and recruit 10 graduate students (annotators) to select the sentences to form the gold standard summary. The annotated 36 bug reports (SDS) are then used as a benchmark in the task of bug report summarization.

---

[4]https://bugs.eclipse.org/bugs/show_bug.cgi?id=214066, last checked July. 1, 2017

Fig. 1. Bug 214066: an example of bug report. A bug report usually consists of a title, some specified items, a description, and some comments.

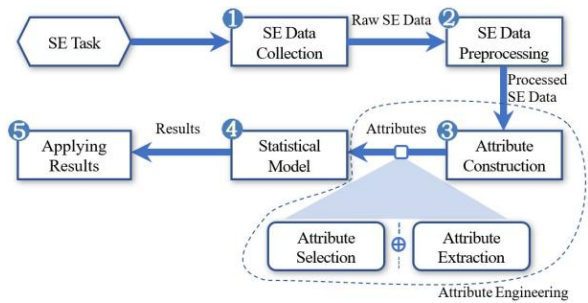

Fig. 2. A general process of supervised approaches in MSR.

In the communities of MSR, one similar task with bug report summarization is bug report enrichment [30], which enriches the description of a bug report with related information from other bug reports. However, the two tasks are different from both the target and usage scenario. For the target, bug report enrichment aims at adding details (e.g., stack traces) to a newly submitted bug report to help developers/triagers fully comprehend it. In contrast, bug report summarization aims at summarizing historical bug reports to help developers/testers quickly get the knowledge in them. For the usage scenario, bug report enrichment enriches the description of a bug report, since most newly submitted bug reports are written in less than 100 words [30]. While bug report summarization summarizes the description and comments of bug reports, when there are too many comments for developers/testers to read in limited time. Hence, bug report summarization is essential to accelerate the daily work of developers/testers.

### B. Attributes in Supervised Approaches

In a supervised approach, attributes are to be constructed and evaluated to train a statistical model. In this subsection, we first review the general workflow of supervised approaches in MSR and then clarify some conceptions related to attributes.

As shown in Fig. 2, the process of supervised approaches in MSR consists of five steps. First, the raw SE data is collected with respect to the task under solving. In the task of bug report summarization, bug reports are collected as the raw SE data [15], [17], [18]. Second, some preprocessing activities (e.g., tokenization, stemming, and stop word removal) can be conducted to clean and format the raw data. Third, a set of attributes are constructed with discriminative information extracted from the preprocessed data. In this step, some related activities, namely attribute selection and attribute extraction, can be optionally conducted to further improve the attributes. Fourth, the values of attributes are evaluated to train a statistical model, e.g., logistic regression, decision tree, etc. Finally, the mining results are applied to solve SE tasks.

The activities related to attributes, e.g., attribute construction, attribute selection, and attribute extraction[5] are vaguely referred to as attribute engineering. For clarity, we present the definitions of these conceptions as follows [31]–[33].

*Attribute Construction* is the process of determining which discriminative information from data should be used to form a set of attributes. These attributes are to be evaluated and fed into statistical models.

*Attribute Selection* is the process of removing redundant and irrelevant attributes from the existing attributes, and selecting relevant attributes. Here, redundant attributes refer to those attributes providing no more information than the selected attributes. In contrast, irrelevant attributes provide no useful information of data.

*Attribute Extraction* is the process of creating new attributes from the functions of the existing attributes, so as to transform the high-dimensional data to a low-dimensional space.

As to these conceptions, attribute selection and attribute extraction could be viewed as the post-processing steps after attribute construction. In MSR, some efforts have been spent on attribute selection and attribute extraction [34]–[37]. In contrast, no research work has been conducted to investigate the methods of attribute construction. Many researchers present new attributes for solving SE tasks without explaining the methods behind constructing these attributes [7], [8]. For example, Sun et al. propose 54 attributes in the task of duplicate bug report detection [8], but they have not explained

---

[5]In the communities of data mining and machine learning, attribute selection and attribute extraction are also known as feature selection and feature extraction.

Q1: *Which methods do you follow to construct these attributes in the paper (the title of the paper)?*

- A: Knowledge Transfer: transfer attributes from similar fields, namely leverage/modify attributes in related work.
- B: Mining Data: manually read the input data of a task and summarize some rules or strategies as attributes.
- C: Heuristic or Experience: identify attributes largely depending on the experience of several researchers (authors) without many reasons.
- D: Crowdsourcing: recruit volunteers to accomplish a task and identify some attributes.
- E: Others.

Q2: *How many persons are involved in the process of attribute construction?*

- A: 1-3 persons
- B: 4-6 persons
- C: 7-10 persons
- D: More than 10 persons

Fig. 3. Questions in the survey.

TABLE I
RESULTS OF THE SURVEY.

why they use *Summary* and *Description* only to construct attributes for bug reports, rather than other items, e.g., *Product*.

In summary, it still remains a challenge for researchers to construct attributes in MSR when tackling a new SE task.

## III. SURVEY OF ATTRIBUTE CONSTRUCTION

Although attributes greatly impact the performance of a supervised approach, no systematical analysis has been conducted in the literature for bug report summarization (and other tasks in MSR). In this section, we conduct a survey to achieve an overview understanding on the current status of attribute construction. In this survey, we seek the research papers related to constructing new attributes in MSR and inquire the authors about the methods they follow in attribute construction and the number of persons involved in this process.

More specifically, we first manually check 166 research papers published in 2012, 2013, and 2014 on the Working Conference on Mining Software Repositories[6], a main international conference focusing on MSR. We find that nearly a quarter of these papers (40 papers) are related to constructing new attributes to automate SE tasks. Then, we send an email to the authors of each attribute-related paper with a survey including two questions (see Fig. 3). In this study, we survey the MSR conference since authors from this conference mostly focus on mining software repositories with newly constructed attributes. An interesting finding is that according to the computer science bibliography DBLP[7], at least one author from 27 out of the 40 surveyed papers has experience in publishing papers from other premier conferences or journals, e.g., ICSE, TSE. Hence, our survey can reveal the methodology on attribute construction not limited to the MSR conference.

Since none of the authors of the 40 papers explicitly presents her/his method of attribute construction, we infer three possible methods, namely knowledge transfer, mining data, and heuristic or experience, from these papers and list them as the first three options in Q1. In addition, we list crowdsourcing as the fourth method to investigate its application status. In case that the authors may have other methods, we also provide the option *E: Others* for the authors to provide their unique methods. In Q2, we analyze the average number of persons involved in attribute construction for a SE task. Such questions aid us in having an intuitive understanding on the methods when researchers construct attributes for a practical SE task.

We receive 20 responses, which means that the response rate is 50%. Even though MSR is a specific research area which limits the number of authors we investigate, the survey still helps us achieve some interesting findings. We present the results of our survey in Table I. We list the selected options of Q1 and Q2 in black. The total number of each selected options is counted in the last row. Since some authors also provide a few comments for their selections, we give some examples of these comments in the last column of Table I. For Q1, we find that these authors pay nearly equal attention to the first three methods and some authors may combine several methods together to construct attributes for an SE task. As to the survey, no author employs crowdsourcing for attribute construction yet. For Q2, the work of attribute construction usually involves a small set of persons, namely 1-3 persons.

Meanwhile these responses also reveal some difficulties of attribute construction in MSR. First, no universal method exists in attribute construction. Sometimes the authors construct new attributes either by an accidental finding in the experiment or an unclear method (see Comments (1) and (4)). Second, the authors tend to try out various attributes in a task, while mostly only 1-3 persons are involved in this process which may limit the attributes that the authors could construct (see Comments (2), (3), and (5)).

Based on the above findings, we propose CA to construct attributes for SE tasks by involving more volunteers.

---

[6]http://www.msrconf.org/ last checked July. 1, 2017
[7]Computer science bibliography DBLP. http://dblp.org/

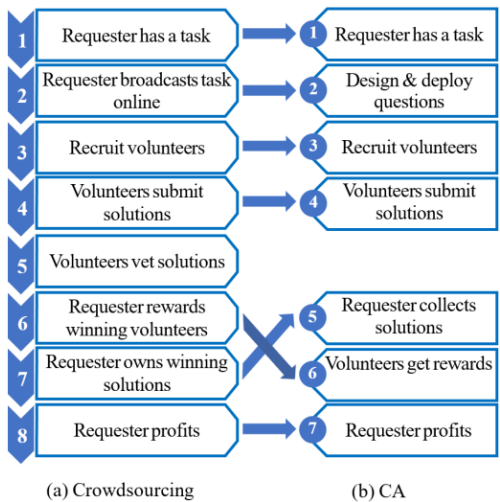

Fig. 4. Roadmaps of crowdsourcing & CA.

---

**Heuristic Construction Rules (HCRs)**

**Phase 1: HCR1**

**Extract candidate attributes from every sentence in the reasons.**

Candidate attributes are selected from the high frequency noun/adjective terms and adjective + noun phrases in the subject or object of a sentence.

**Phase 2: HCR2**

**Remove meaningless candidate attributes.**

(a) Group candidate attributes by synonyms and morphology.

(b) We delete a group, if its candidate attributes do not satisfy any of the following rules:

- The candidate attributes in the group appear in some predefined items of SE data. For a bug report, the typical predefined items include product, priority (importance), status, component, reporter name, commenter name, etc.

- The candidate attributes can be transformed into some measurements. For example, "similar with, similarity, related to, duplicate" are related to the similarity, "solution to the problem, solving the problem, suggestion to the bug" express the relationships between two parts, and "rich in content, (sentence is) simple, concrete advice" show the complexity/importance of a sentence, etc.

**Phase 3: HCR3**

**Calculate and merge candidate attributes to achieve new attributes.**

(a) We use criteria to measure the candidate attributes that are the same with attributes in related work, are predefined items or keywords, are related to some measurements, and can be transformed into some measurements.

(b) We merge the candidate attributes,

- if the inputs of the calculation metrics (measurements) are the same,

- and the calculation metrics output the same values for all the inputs.

Fig. 5. Heuristic construction rules.

---

## IV. Crowd-Attribute and Crowdsourcing SE Platform

In this section, we first present the roadmap of CA, then briefly introduce our new tool CSEP to facilitate CA[8].

### A. Roadmap of Crowd-Attribute

Since being proposed in 2005, crowdsourcing has been employed to solve a wide range of tasks [38]. As to [39], a general process of crowdsourcing consists of eight steps (see Fig. 4(a)). When a requester has a task for solving, she/he broadcasts it online and calls for some online volunteers to propose their solutions. The volunteers submit their solutions and vet others' solutions. After the vetting step, the volunteers with the most votes may win out and be rewarded by the requester. Then, the requester exploits the final solution to solve the task. Recently, a lot of research work [29], [40], [41] employ crowdsourcing to facilitate SE tasks. However, no related work has been performed on constructing attributes.

Following the general steps of crowdsourcing, we take the SE task of bug report summarization as an example to explain the roadmap of CA (see Fig. 4(b)). In CA, a requester first determines the task for solving, i.e., constructing new attributes for bug report summarization. Second, the requester broadcasts the crowdsourcing task online, namely designing a set of task-related questions for the volunteers and choosing a platform to deploy these questions. In this example, the requester asks the volunteers for the reasons on how they determine the summary sentences in a bug report. The requester broadcasts the questions on CSEP. Third, the volunteers are recruited to answer the questions. In such a way, a solution in our study refers to the answers proposed by a volunteer to the questions. Fourth, every volunteer submits her/his solution, i.e., her/his answers to the questions. Since we need to construct attributes from the solutions, all the solutions are collected and examined without a vetting process. After examining the solutions, the volunteers providing high quality solutions are rewarded with a gift, e.g., a USB flash disk in this study. Finally, the requester profits by constructing new attributes under the inspiration of the solutions, e.g., the reasons to select summary sentences.

In this study, we design Heuristic Construction Rules (HCRs) in Fig. 5 to guide the action of attribute construction from volunteers' reasons. HCRs consist of three phases.

In HCR1, some candidate attributes are extracted from every reason. Specifically, we analyze the part-of-speech and the structure of each reason. The subject and object of the reasons are identified. Since attributes are usually expressed as nouns or adjectives, e.g., "the length of the sentence" or "the sentence is long", we check whether the subject or object of each reason is a noun/adjective term or adjective+noun phrase. If it is true, we take the term or phrase as a candidate attribute. HCR1 limits requesters to select only one candidate attribute from each reason. Hence, requesters select the candidate attribute according to the frequency of terms/phrases counted by all the reasons. If no candidate attribute is identified, e.g., it is not a complete sentence, we remove the reason. We make the candidate attribute and its corresponding reason as a candidate pair, i.e., <candidate attribute, reason>. In this phase, a set of candidate pairs are collected.

---

[8]Supplemental materials on suggestions of CA and source code of CSEP are available at http://oscar-lab.org/people/%7excli/open/crowdsourcing/

In HCR2, some meaningless candidate attributes are detected and removed. First, we group the candidate pairs according to the synonyms and morphology of candidate attributes. Thus, pairs of "<the length, ...>" and "<lengths, ...>" are put in the same group. Second, for each group, we analyze the meaning of the candidate attributes. We reserve a group, if the candidate attributes in the group appear in some predefined items of SE data, e.g., the items *Product* and *Priority* in a bug report, since these items are more discriminative than free text (e.g., comments in a bug report). We also reserve candidate attributes that can be transformed into some measurements, e.g., "length of text", "frequency of terms", and "similarity" (similar), which are widely used in the literature. Otherwise, a group of candidate attributes is removed. At the last of this phase, we select one of the candidate attributes in each reserved group as a representative, since all the candidate attributes in the same group are synonyms of the same meaning.

Finally, HCR3 calculates the values of the representative candidate attributes and merges candidate attributes with the same calculation metrics to achieve the final attributes. In the calculation step, all candidate attributes in the same group are calculated by the same criterion:

- C1. For candidate attributes that are the same with the attributes in the related work, the same methods are followed to calculate the attributes.

- C2. If the candidate attribute is a predefined item or a keyword of SE data, we enumerate the status of the item or keyword as attribute values, e.g., using 0 and 1 to denote whether a sentence related to an item, and using 0, 1, 2 to denote the bug report *Priority* P1, P2, and P3.

- C3. For attributes related to some measurements, the mathematic definitions of the measurements can be followed to calculate the corresponding information in SE data, e.g., calculating the length or similarity of sentences.

- C4. For the remaining attributes, we try to transform them into some measurements, e.g., transforming a candidate attribute into a type of length or similarity.

If no criterion is satisfied, the candidate attribute is deleted. In the merging step, we merge two candidate attributes when they use the same metrics for calculation. Two calculation metrics are the same, if the inputs of the calculation metrics are the same and they output the same values for all the inputs. For example, "<the length, ...>" and "<long, ...>" can be merged, if requesters calculate both candidate attributes by the number of characters in a sentence. After merging, the final attributes are achieved.

With these heuristic rules, the attributes could be constructed step by step. Some suggestion we shall demonstrate the application of HCRs to bug report summarization in Section V.

As to its roadmap, CA is far different from two existing methods, namely knowledge transfer and heuristic or experience. In contrast, CA is somehow similar as the method of mining data, since both methods investigate SE data to construct new attributes. The key difference between CA and mining data is that CA is more likely to construct more effective attributes than mining data, because in the process of attribute construction, a requester (researcher) in CA is inspired by many volunteers rather than by herself/himself only in the method of mining data.

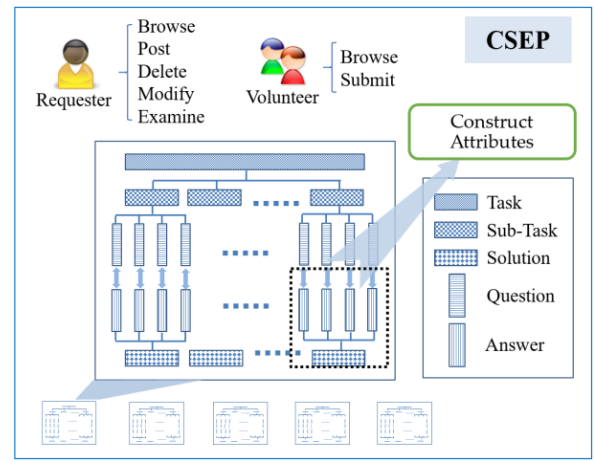

Fig. 6. Roles and functions of CSEP.

### B. Crowdsourcing SE Platform

To facilitate the process of CA, we implement a new crowdsourcing based tool CSEP. CSEP is implemented in J2EE using MySQL Server 5.5, compiled with MyEclipse 8.5, and deployed with Tomcat 6.0.

As shown in Fig. 6, there are two roles (volunteer and requester) in CSEP. A volunteer can browse existing tasks, join tasks, and submit solutions. A requester can post, modify, and delete tasks. In addition, a requester can examine the solutions by volunteers to construct new attributes for SE tasks. In CSEP, a crowdsourcing task usually consists of several subtasks and each subtask is associated with a group of questions predefined by a requester. In such a way, a volunteer's solution is composed of all her/his answers to the questions.

After a volunteer logins into CSEP, she/he can find a list of existing tasks under solving (see Fig. 7(A)). If the volunteer joins any task, CSEP will explain the related conceptions of the task for the volunteers. After a volunteer joins a task, she/he can submit her/his solution. For example, in the task of bug report summarization, a volunteer is assigned several subtasks. In each subtask, the volunteer is presented a bug report with several questions (see Fig. 7(B)), including which sentences are selected as the summary and why she/he makes this decision. The volunteer submits the answers to the questions as her/his solution. In CSEP, all the bug reports can be selected in random or a predefined order. Meanwhile, CSEP can decide the initial maximum times to present a bug report by counting the presented times of each bug report. If the presented times of all the bug reports reach this initial setting, CSEP resets the counter to continue selecting bug reports. CSEP records every operation of the volunteers to ensure that no volunteer will receive the same bug report twice.

After submitting all the solutions, a requester can directly browse the solutions and construct attributes by the interface of CSEP in Fig. 7(C). In this platform, the "AnswerComment" and "AnswerAttribute" columns are editable. The "Answer-Comment" column is designed for requesters to record the candidate attributes of a sentence, the candidate attribute group IDs, their understanding of the sentence, the possible calculation metrics for attributes, the abbreviation for attributes, etc. The contents of "AnswerComment" are changed by requesters at different phases of HCRs. The "AnswerAttribute" column is only used to record the final attributes.

Fig. 7. CSEP screenshots: (A) Volunteer browses tasks. (B) Volunteer submits solutions. (C) Requester constructs attributes.

## V. NEW ATTRIBUTES AND LRCA FOR BUG REPORT SUMMARIZATION

In this section, we present how to achieve new attributes by applying CA to bug report summarization and introduce the new supervised approach LRCA.

### A. Applying CA to Bug Report Summarization

This subsection applies CA to bug report summarization.

*1) Requester Has A Task:* In this study, we aim to construct effective attributes for bug report summarization. We download the 'fixed' bug reports till Dec. 31st, 2015 from bug repositories of Mozilla, Eclipse, KDE, and Gnome as the data set for further processing.

*2) Design & Deploy Questions:* The crowdsourcing task is composed of several subtasks. For each subtask, a bug report and two questions are to be specified before they are deployed. In this step, we first prepare a set of crowdsourcing bug reports. Since the volunteers may lack experience in bug report summarization, we define two criteria for preparing the crowdsourcing bug reports as follows:

*Criterion* 1: We prefer bug reports without long stack traces.

*Criterion* 2: We prefer bug reports with fewer technical abbreviations and chunks of code.

These criteria are similar as in [15]. With the above criteria, we iteratively choose the crowdsourcing bug reports from the downloaded data set. More specifically, we randomly check 10 bug reports from the data set each time. Out of these bug reports, one best satisfying the above criteria is kept as the

**Q3: Do you have difficulty with comprehending the above bug reports?**
- 5: Very easy to understand.
- 4: Easy to understand.
- 3: I am not sure.
- 2: Hard to understand.
- 1: Very hard to understand.

**Q4: Please choose the summary sentences and provide the reasons for each summary sentence.**

Fig. 8. Questions for bug report summarization.

crowdsourcing bug reports. We iterate this process until we collect enough crowdsourcing bug reports. In this case study, we prepare 30 crowdsourcing bug reports.

Second, we design two questions (Q3 and Q4 in Fig. 8) for each subtask. We check in Q3 whether the volunteers could well understand the bug reports. In Q4, we ask the volunteers to select sentences as the summary and write their reasons.

Finally, we deploy the crowdsourcing bug reports and related questions on CSEP. Initially, the initial maximum time to present a bug report is set to two. This setting guarantees that each crowdsourcing bug report can be viewed by at least two volunteers based on the recruited volunteers number as mentioned in the next step. Every volunteer is requested to investigate three randomly chosen bug reports in two weeks.

*3) Recruit Volunteers:* We send an email to 450 students in the School of Software, Dalian University of Technology. In this email, we invite them to participate in this crowdsourcing task with a reward of a USB flash disk worth $4. This task successfully attracts 21 volunteers, i.e., the participation rate is 4.67%. In some crowdsourcing platforms such as Mechanical Turk, the requesters sometimes utilize a qualification test to evaluate the background of volunteers [42], [43]. In this study, since all the volunteers have English and computer science background, no qualification test is employed.

*4) Volunteers Submit Solutions:* The 21 volunteers login into CSEP and submit their solutions. On average, each

TABLE II
Examples of volunteers reasons and constructed attributes.

| Selected Summary Sentence | Volunteer's Reason | Requester's Comment | New Attribute |
|---|---|---|---|
| S1: The idea is that if one prefers to use a web based e-mail account, they should have the option to configure firefox to log into that web based e-mail account and open a com-pose new message window. | R1: the sentence is **long** and the writer puts forward his idea. | long: the length of a sentence may be an attribute | SLEN |
| | R2: **a rich content** | rich content: we can measure rich content with the sentence length | |
| S2: Being able to see and sort the history. | R3: **reporter** provides some suggestions | reporter: we can regard the reporter as an attribute | REP |
| S3: Please add the Mozilla history window in Firefox! | R4: It is **similar with** the sentences in the **bug report**. | similar with bug report: calculate similarity | SWD |
| | R5: **reporter** provides some suggestions | reporter: be an attribute | REP |

TABLE III
Basic information for attribute construction.

| Req | HCR1 | HCR2 | | HCR3 | | | |
|---|---|---|---|---|---|---|---|
| | #Pair | #Group | #Reserved | #C1 | #C2 | #C3 | #C4 |
| Req1 | 197 | 45 | 42 | 16 | 11 | 10 | 5 |
| Req2 | 190 | 37 | 37 | 14 | 9 | 6 | 8 |

volunteer spends 25.9 minutes on this task.

*5) Requester Collects Answers:* With CSEP, we successfully collect 21 solutions. Since a bug report contains a few sentences, we collect 332 reasons for Q4 in total.

*6) Volunteers Get Rewards:* We award the volunteers who provide rational solutions. Out of the 21 volunteers, 19 volunteers are awarded a USB flash disk, since two volunteers provide no reasons for Q4.

*7) Requester Profits:* In this step, we achieve new attributes by conducting two actions sequentially, namely answer filtering and attribute construction.

In the action of answer filtering, we compare the answers to Q4 of each volunteer. If a sentence is viewed by more than one volunteer and all the volunteers fill reasons with the consistent summary sentence decision, we reserve the answers related to this sentence. Out of the 332 answers for Q4, 200 answers are finally reserved, i.e., the reserved rate is 60.24%. The reserved rate is reasonable compared with some similar human annotating processes [15].

In the action of attribute construction, new attributes are constructed under the inspiration of these 200 reasons with HCRs. In Table II, we present some examples of the volunteers' reasons and related new attributes. The first two columns are the summary sentences selected by the volunteers and their reasons. The remaining columns present some requester's comments and newly constructed attributes. Table III presents the basic information for the attribute construction process, including the number of candidate attribute pairs in HCR1, the number of groups and reserved groups in HCR2, and the number of groups applying HCR3-C1 to HCR3-C4. In this study, the second and third authors act as the requesters (denote as Req1 and Req2) to construct attributes independently.

First, we apply HCR1 to extract some candidate attributes from the reasons. The two requesters find 197 and 190 candidate attribute pairs from the reasons. For example, they extract 'long' as a candidate attribute from R1 in Table II, since it is an adjective term and frequently appears in other reasons (for brevity, some other reasons are not shown). Similarly, in Table II, "a rich content" is the only phrase in R2. The adjective term "similar" is the object of sentence R4. The noun term "reporter" is the subject of sentences R3 and R5. We label these candidate attributes in bold in Table II.

Second, we apply HCR2 to remove some meaningless groups of candidate attributes. In this process, the two requesters classify these pairs into 45 and 37 groups according to the synonyms and morphology of candidate attributes, e.g., <reporter, R3> and <reporter, R5> are put in the same group. Then, Req1 removes three meaningless groups. In contrast, all the groups are reserved by Req2. For the examples in Table II, the candidate attribute "reporter" is a predefined item of bug reports. Terms "long" and "similar" are some measurements of bug reports. Meanwhile, "rich content" is transformed into some measurements related to "long" by Req1.

Finally, we apply HCR3 to achieve final attributes by calculating and merging similar groups of candidate attributes.

In the calculation step, requesters may follow different criteria to calculate candidate attributes. For example, Req1 finds that 16 groups of candidate attributes already exist in previous studies for bug report summarization and follows HCR3-C1 for calculation. In contrast, the number of groups applied HCR3-C1 is 14 for Req2. According to Table III, we find that many groups of candidate attributes can be easily calculated by following previous studies (HCR3-C1) or using enumeration metrics (HCR3-C2). For the candidate attributes "long" and "rich content", requesters follow previous studies [15] to calculate the number of characters in the sentences. The candidate attribute "reporter" is a bug report item which can be represented as 0 or 1 for different sentences by HCR3-C2. For the candidate attribute "similar", both requesters employ Vector Space Model (VSM) [44] for calculation.

In the merging step, requesters merge similar groups of candidate attributes. For example, the groups related to 'long' and 'rich content' are merged, which evaluates the length of a sentence. Eventually, according to reasons R1-R5 in Table II, we construct three attributes, namely SLEN, REP, and SWD.

Out of the 200 reasons, the two authors can construct the same attributes from 153 reasons. The concordance rate is 76.5 %. The two authors conduct a pair-wise discussion to resolve their conflicts. At last, 11 attributes are successfully

| # | Attribute | Short description |
|---|-----------|-------------------|
| 1 | SWT | Similarity With the Topic of bug report |
| 2 | SWD | Similarity With Description |
| 3 | DUP | is the sentence a DUPlicate sentence |
| 4 | SLEN | the normalized Sentence LENgth |
| 5 | SI | the normalized Sentence Importance |
| 6 | SLOC | the normalized Sentence LOCation |
| 7 | CLEN | the normalized Comment LENgth |
| 8 | DES | is the sentence in the DEScription |
| 9 | CCW | does the sentence Contain Certain Words |
| 10 | CODE | is the sentence in a piece of CODE snippet |
| 11 | REP | is the sentence written by the REPorter |

constructed for bug report summarization.

In this process, we also achieve an interesting finding that two volunteers may choose the same summary sentence with distinct reasons. For example, one volunteer selects the sentence S3 in Table II under the consideration that 'R5: reporter provides some suggestions', while another volunteer selects it due to the reason that 'R4: It is similar with the sentences in the bug report'. It confirms that involving more volunteers could provide more ideas from distinct aspects.

### B. New Attributes and LRCA

In this subsection, we first explain these new attributes achieved by CA. Then, we outline the new supervised algorithm LRCA for summarizing bug reports.

Table IV provides a short description of the 11 attributes. Since some attributes in the table rely on the similarity, to generalize the study and reduce the influence of requesters' background knowledge, we only employ Vector Space Model (VSM) [44], a widely used model in information retrieval, to evaluate the similarity between two text units (e.g., sentences, comments, descriptions). In VSM, every text unit is represented as a vector, where each dimension corresponds to the Term Frequency-Inverse Document Frequency (TF-IDF) value of a term in this text unit. In information retrieval, the TF-IDF value can reflect the importance of a term to a text unit in a corpus [45]. In this study, for each term $t$ in a text unit $d$, the TF-IDF value calculates the term weight as follows:

$$\mathrm{TF\text{-}IDF}_{t,d} = f_{t,d} * \log \frac{N}{n_t}, \quad (1)$$

where $f_{t,d}$ denotes the number of times that $t$ occurs in $d$, $n_t$ denotes the number of text units containing $t$, and $N$ denotes the number of text units in the corpus (e.g., SDS). Given two text units with their vectors of TF-IDF values, VSM evaluates the similarity between the text units by calculating the cosine of the angle between the vectors.

Given a sentence $s$ in a bug report $T$, we detail the 11 attributes as follows:

· SWT measures the similarity between the sentence $s$ and the bug report $T$ with VSM.

[Reason] Some volunteers select summary sentences by comparing the sentences with the topic of the entire bug reports, e.g., "it is related to the topic", "it contains some key words related to the topic". Hence, requesters transform such reasons into the relatedness between a sentence and the bug report topic by HCR3-C4, and then use VSM to calculate the "relatedness". However, we notice that "topic" may have diverse definitions. Previous studies basically extract keywords in a document as the topic [46] or transform all the words into a low-dimension vectors to represent the topic [47], e.g., LDA, PLSA. Since words are the basic elements for these methods, in this study, requesters utilize words in the entire bug reports to represent the bug report topic to minimize the influence of sophisticated measurement methods.

· SWD measures the similarity between the sentence $s$ and the description in the bug report $T$ with VSM. If $s$ belongs to the description, its SWD is set to 1.

[Reason] This attribute is elicited by the reasons like "it is similar with the sentences in the description", "this sentence explains the best way of solving the problem", etc. For the first sentence, both requesters follow HCR3-C3 to calculate the similarity between a sentence and the bug report description. For the second sentence, Req2 regards the object "solving the problem" as the candidate attribute. He takes the "problem" as the bug report description, since bug report description usually illustrates the problem of a bug. Then he translates the candidate attribute as a type of similarity to evaluate whether a sentence is related to the "problem" by HCR3-C4.

· DUP checks if the sentence $s$ is a duplicate of another sentence $Y$ located before $s$. If $s$ is a duplicate of $Y$, the DUP of $s$ is set to 1, otherwise 0. Here, $s$ is said to be a duplicate of $Y$ if the similarity between $s$ and $Y$ is larger than a predefined threshold. In this study, the threshold is set to 0.8 after parameter tuning.

[Reason] As to the reasons by volunteers, the sentence $Y$ is more likely to be chosen as a summary sentence rather than the duplicate sentence $s$. To detect duplicate sentences, requesters translate "duplication" as a type of similarity for calculation.

· SLEN measures the length of the sentence $s$, normalized by the maximum length of sentences in the bug report $T$. Here, the length of a sentence is defined as the number of characters in this sentence.

[Reason] SLEN is identified by reasons R1 and R2 in Table II. As an attribute in the previous study [15], requesters utilize HCR3-C1 to calculate this attribute.

· *SI measures the importance of a sentence.* For the sentence $s$, we add up the TF-IDF value of each term in $s$ to represent its importance. SI returns the importance of the sentence $s$ normalized by the maximum importance value in the bug report $T$.

[Reason] SI is an attribute constructed by HCR3-C4. This attribute is inspired by the reasons such as "this sentence contains many special words like createWidget and readAndDispatch". Requesters select the adjective+noun phrase "special words" as candidate attributes according to HCR1. To calculate the values of these special words, requesters still follow the widely used Vector Space Model. The model naturally calculates the importance of a word by its TF-IDF value.

· SLOC measures the location of the sentence $s$, normalized by the number of sentences in this bug report. If $s$ is the third

sentence of a bug report with ten sentences, SLOC of $s$ is 0.3.

[Reason] SLOC is an attribute that can be calculated in the same way as previous studies [15] according to HCR3-C1.

· *CLEN* measures the length of the description/comment containing the sentence $s$, normalized by the maximum length of the description/comments in the bug report $T$. Here, the length of a description/comment is defined as the number of characters contained in this description/comment.

[Reason] CLEN is an attribute constructed by HCR3-C3. The "length" is calculated in the same way as SLEN.

· *DES* indicates whether the sentence $s$ is in the description of the bug report $T$. If $s$ is contained in the description, its DES is set to 1, otherwise 0.

[Reason] This attribute is constructed since some volunteers prefer "selecting sentences of bug report description". In the reasons, "bug report description" is a predefined item of a bug report. Requesters follow HCR3-C2 to calculate its value.

· *CCW* indicates whether the sentence $s$ provides a hyper-link address or contains a key term "problem". If the sentence $s$ provides a hyperlink address, its CCW is set to 0. On the other hand, if $s$ contains the term "problem", its CCW is set to 1. Otherwise, its CCW is set to 0.5.

[Reason] As to the reasons by volunteers, a sentence providing a hyperlink address is unlikely to be a summary sentence. On the contrary, a sentence containing the key term "problem" usually provides either the root cause or the phenomenon of the related bug. Hence, such a sentence is likely to be one of the summary sentences. Requesters utilize HCR3-C2 to enumerate the keywords as 0, 0.5, 1 after being normalized.

· *CODE* indicates whether the sentence $s$ is in a piece of code snippet. Its CODE is set to 1 if $s$ is in a piece of code snippet, otherwise 0. In this study, we detect code snippets with a set of heuristic patterns:

1) It starts with 'db2', 'proc', 'public', '>', '/*', '//'.

2) It contains '<', 'if.*(.*', 'sql', '{', '}', 'public static', and '='.

3) It ends with ';'.

[Reason] Since volunteers comment some sentences as "there is nothing but codes", requesters take source codes as special keywords in bug reports. According to HCR3-C2, the attribute is mapped into 0 or 1 by detecting source codes with heuristic patterns.

· *REP* is set to 1, if the sentence $s$ is provided by the reporter of the bug report $T$, otherwise 0.

[Reason] REP is identified by the candidate attributes of R3 and R5 in Table II. Requesters directly follow the previous study [15] to calculate this attribute according to HCR3-C1.

With these new attributes, we outline our new supervised algorithm LRCA for summarizing bug reports. To demonstrate the effectiveness of the attributes constructed by CA, we follow the BRC framework to design the LRCA algorithm. There are two main steps in the BRC framework, namely the training step and the testing step [15]. In the training step, BRC inputs a set of labeled bug reports. Each sentence in the bug reports is labeled as 1 or 0 which means a summary sentence or an ordinary sentence respectively. These sentences are used to calculate the 24 attributes transferred from generic conversation-based summarization. As a result, each sentence is represented as a vector of 24 dimensions. BRC inputs these vectors and the corresponding labels into a logistic regression model [48] to generate a statistical model for bug report summarization. In the testing step, BRC transforms the sentences in a new bug report into similar vectors, and feeds the vectors into the statistical model to get the probability values that these sentences belong to the summary sentence set. BRC selects the top ranked sentences with the highest probability values as the summary sentences of the bug report.

In contrast to BRC, LRCA transforms the sentences in each bug report according to the 11 attributes constructed by CA, and trains and tests the logical regression statical model with vectors of 11 dimensions. Since BRC and LRCA follow the same framework to summarize bug reports, if LRCA constantly outperforms BRC, it means that the attributes from CA are more effective than the transferred attributes by BRC.

## VI. EXPERIMENTS

This section presents the Research Questions (RQs), the experiment setup, the baseline algorithms, the data set and metrics, and the answers to these RQs.

### A. Research Questions

**RQ1**: *Can CA construct meaningful attributes?*

This RQ investigates whether CA can construct domain specific attributes with low correlations.

**RQ2**: *Can LRCA improve the comparative approaches?*

In this RQ, we evaluate the effectiveness of attributes by CA for bug report summarization. We compare LRCA over the publicly available data set, namely SDS, against the existing approaches, including supervised and unsupervised ones.

**RQ3**: *How does the number of volunteers influence the effectiveness of LRCA?*

CA in this study involves a number of volunteers. This RQ evaluates the influence of the number of volunteers. More specifically, we will achieve the attributes contributed by distinct sized combinations of volunteers in CA and evaluate the performance of LRCA with these attributes.

**RQ4**: *Can LRCA perform well over large scale data sets for bug report summarization?*

In this RQ, we build a series of large scale data sets for bug report summarization and evaluate the performance of LRCA against the existing approaches.

**RQ5**: *Can we employ interested volunteers with the necessary knowledge to participate in CA?*

We investigate in this RQ whether there exist interested volunteers with the sufficient knowledge to participate in CA.

### B. Experiment Setup

In the experiments, all the algorithms are implemented in Java JDK1.8.0 31, and run on PCs running 64-bit Win 7 with Intel Core(TM) i5-3470 CPU and 8G memory.

For a fair comparison, we adopt the same settings as in the literature [15], [17] in all the algorithms. Every algorithm ranks the sentences in a bug report by either their predicted probabilities [15] or the predefined measure values [17], [18] in descending order. Then the top ranked sentences are selected out, one by one, to form the summary until the number of words in the summary reaches a predefined threshold, namely 25% of the total number of words in this bug report.

TABLE V
ATTRIBUTES FOR BASELINE ALGORITHMS.

| # | Algorithm | Attribute | Short description | LRCA Attr. |
|---|-----------|-----------|------------------|-----------|
| 1 | | MXS | max $Sprob$ score | |
| 2 | | MNS | Mean $Sprob$ score | |
| 3 | | SMS | Sum of $Sprob$ score | |
| 4 | | MXT | Max $Tprob$ score | |
| 5 | | MNT | Mean $Tprob$ score | |
| 6 | | SMT | Sum of $Tprob$ score | |
| 7 | | TLOC | Position in turn | |
| 8 | | CLOC | Position in conv. | SLOC |
| 9 | | SLEN | Word count, globally normalized | SLEN |
| 10 | | SLEN2 | Word count, locally normalized | |
| 11 | | TPOS1 | Time from beg. of conv. to turn | |
| 12 | BRC | TPOS2 | Time from turn to end of conv. | |
| 13 | | PPAU | Time btwn. current and prior turn | |
| 14 | | SPAU | Time btwn. current and next turn | |
| 15 | | COS1 | Cos. of conv. splits, w/ $Sprob$ | |
| 16 | | COS2 | Cos. of conv. splits, w/ $Tprob$ | |
| 17 | | CENT1 | Cos. of sentence & conv., w/ $Sprob$ | |
| 18 | | CENT2 | Cos. of sentence & conv., w/ $Tprob$ | |
| 19 | | PENT | Entropy of conv. up to sentence | |
| 20 | | SENT | Entropy of conv. after the sentence | |
| 21 | | THISENT | Entropy of current sentence | |
| 22 | | DOM | Participator dominance in words | |
| 23 | | BEGAUTH | Is first participator (0/1) | REP |
| 24 | | CWS | Negative ClueWordScore | |
| 25 | Centroid MMR | WORD | TF-IDF value of each word | |
| 26 | Grasshopper DivRank | SENTENCE | Simi. btwn. a sentence and every sentence in the bug report | |
| 27 | | TITLE | Simi. btwn. the sentence and title | |
| 28 | Hurried | DES | Is the sentence in the description | DES |
| 29 | | SENTIMENT | The sentiment of a sentence | |

### C. Baseline Algorithms

We detail the baseline algorithms in this subsection. To have a clear understanding of the baselines, Table V summarizes the attributes used in these algorithms as well as a short description of each attribute. In the last column, we label the attributes in LRCA that are the same with previous attributes.

*1) Bug Report Classifier (BRC):* BRC [15] is a supervised algorithm to summarize bug reports with 24 conversation-based attributes (Attributes #1 – #24). The first six attributes refer to $Sprob$ and $Tprob$. $Sprob$ is the probability that a word belongs to a participator. If word $w$ occurs 10 times in a bug report and participator $p$ uses $w$ for 3 times, then $Sprob$ of $w$ for $p$ is 0.3. Similarly, $Tprob$ is the probability that a word belongs to a comment or the description. The first six attributes measure the max, mean, and sum of $Sprob$ and $Tprob$ of a sentence. Attributes #7 to #14 measure the position (TLOC, CLOC), length (SLEN, SLEN2), and submitted time (TPOS1, TPOS2, PPAU, SPAU) of a sentence under different metrics. Attributes #15 to #21 analyze the semantic changes between sentences or comments. Specifically, COS1 and COS2 measure the semantic relatedness before and after the current comment. CENT1 and CENT2 measure the semantic relatedness between a sentence and the related comment. PENT, SENT, and THISENT calculate the informativeness (entropy) of different comments. Attributes #22 and #23 are related to the participators. DOM detects the participator that writes most words and BEGAUTH detects the first participator (reporter). At last, CWS detects the unique words of a sentence.

*2) Centroid:* Mani et al. [18] augment four unsupervised algorithms Centroid, MMR, Grasshopper and DivRank with a preprocessing step for bug report summarization. Centroid transforms sentences in a bug report into word-based vectors. Each entry of a vector is the TF-IDF value of a word. The vector size is the vocabulary of the bug report. Thus, words are the attributes of Centroid (Attribute #25). Centroid averages the word-based vectors to form a pseudo vector and selects the summary sentences according to the cosine similarity between the word-based vector and the pseudo vector.

*3) Maximal Marginal Relevance (MMR):* MMR constructs the same word-based vectors and pseudo vector as Centroid. Based on these word-based attributes (Attribute #25), a summary sentence is selected if it is similar with the pseudo vector and dissimilar with the previously selected sentences. Dissimilarity is calculated as the negative of cosine similarity.

*4) Graph Random-walk with Absorbing StateS that HOPs among PEaks for Ranking (Grasshopper):* Grasshopper constructs a sentence graph of the new bug reports for summarization. In the graph, each vertex is a sentence and each edge is the cosine similarity of two sentences. Then a random walk is conducted on the graph. Grasshopper selects sentences by counting the visited numbers of these sentences. Since the basic elements of Grasshopper are the sentence similarity, the attributes of Grasshopper can be regarded as the similarity between a sentence and every sentence in the bug report (Attribute #26). The attribute number is the sentence number.

*5) Diverse Rank (DivRank):* DivRank is an improved algorithm of Grasshopper. It enhances the random walk process by considering not only the similarity between sentences, but also the previously visited numbers of the sentences. Hence, it shares the same attributes with Grasshopper (Attribute #26).

*6) Hurried:* Lotufo et al. [17] manually mine three attributes (#27–#29) for unsupervised bug report summarization. Hurried utilizes PageRank [49] to select a summary sentence with consideration of its similarity with the bug report title, the sentences in the bug report description (the attribute DES), and the sentiment of the sentences.

Besides the supervised algorithm BRC, we compare LRCA with five unsupervised algorithms, since all these algorithms target bug report summarization and they also have attribute-related elements. Obviously, the attribute-related elements in the unsupervised algorithms are different from the 11 attributes in LRCA. For Centroid, MMR, Grasshopper, and DivRank, they only use words or sentences as attributes. In contrast, the 11 attributes elicited by volunteers are quite different and more explainable. For Hurried, the only overlapped attribute with LRCA is DES. However, LRCA uses DES based on the fact that volunteers select summary sentences in consideration of the description. In addition, we note that there are some overlapped attributes between BRC and LRCA, e.g., REP, SLOC (named as CLOC in BRC), etc. Since both the two algorithms consider many attributes, we conduct a thorough analysis of these attributes in Section VII-A

## D. Data set and Metrics

In RQ2 and RQ3, we compare the performance of algorithms over the data set SDS.

To the best of our knowledge, SDS is the only widely-used public data set for bug report summarization with 36 annotated bug reports of 2,361 sentences. For each unsupervised approach, all sentences in a bug report are evaluated and sorted according to their measure values to form the summary. In contrast, in the existing supervised approach BRC [15], Rastkar et al. employ a leave-one-out framework to summarize bug reports. More specifically, given 36 bug reports in SDS, each bug report is selected out as the new bug report for summarization and the remaining bug reports are used as its training set. In this study, LRCA follows the leave-one-out framework to train models.

In the literature [15], [17], [18], four metrics are employed to evaluate algorithms, namely *Precision*, *Recall*, *F-score*, and *Pyramid*. The four metrics are defined as follows.

$$Precision = Num_{success}/Num_{selected}, \qquad (2)$$

where $Num_{success}$ is the number of selected summary sentences which belong to the gold standard summary and $Num_{selected}$ is the number of selected summary sentences by an algorithm. *Precision* measures the portion of selected summary sentences which belong to the gold standard summary.

$$Recall = Num_{success}/Num_{total}, \qquad (3)$$

where $Num_{total}$ is the total number of summary sentences in the gold standard summary. *Recall* can be interpreted as the portion of sentences from the gold standard summary which are selected by an algorithm.

$$F\text{-}score = \frac{2 * Precision * Recall}{Precision + Recall}, \qquad (4)$$

*F-score* indicates an overall performance of an algorithm, which is a weighted average of *Precision* and *Recall*.

$$Pyramid = Num_{total-links}/Num_{max-links}, \qquad (5)$$

*Pyramid* measures the summarization from the perspective of annotators. $Num_{total-links}$ represents the total number of times that the sentences in the summary are voted by the annotators, while $Num_{max-links}$ represents the maximum possible total votes for that summary length.

## E. Answer to RQs

*1) Answer to RQ1:* In this RQ, we investigate whether CA can construct meaningful attributes. Meaningful attributes mean that the attributes may carry the domain knowledge in the bug reports and they are not correlated to each other.

From the 11 identified attributes, we find that CA can find both the generic conversation-based attributes and domain specific attributes. Since bug reports are a special type of conversation, CA constructs several common attributes as for the generic conversation-based summarization, e.g., SWT, DUP, and SLEN. Meanwhile, many domain specific attributes can also be constructed by CA. For example, since the description of a bug report may describe the problem of a bug, it usually attracts more attention from the participators in a bug report than in generic conversations, e.g., meeting conversations. CA constructs two attributes regarding to the description of a bug report, namely SWD and DES. Another difference is that several certain words indicating the summary sentences, e.g., the word 'problem', seldom occur in a generic conversation. The volunteers can inspire the requesters to find these domain specific words. At last, the volunteers also find that CODE is an important attribute in bug reports. In contrast, code snippets may never appear in a generic conversation. Hence, CA has the ability to capture both generic conversation-based attributes and domain specific attributes.

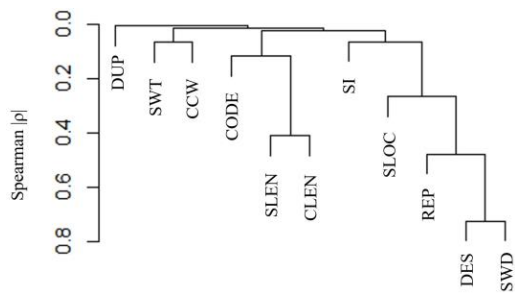

Fig. 9. Correlation analysis on the attributes.

In addition to the qualitative analysis, we also check the pair pairwise correlation between the attributes constructed by CA. The correlation is measured by Spearman rank correlation test ($\rho$), a widely used statistics method which is robust to non-normally distributed data [50], [51]. Given two attributes, Spearman rank correlation test ranks the values of each attribute calculated on a set of instances, e.g., the 2,361 sentences in SDS data set. Then it compares the relative position of each instance in the two ranking lists to calculate the correlation of the two attributes. If two attributes are orthogonality, the value of $\rho$ is 0. If they are highly correlated, the value of $\rho$ is -1 or +1. The value of $\rho$ is between -1 to +1. Fig. 9 presents the results of Spearman hierarchical clustering on the 11 attributes. The results are calculated by the "varclus" function in the R package of "Hmisc". We find that no correlation between two attributes exceeds 0.8 and the correlation between most attributes is lower than 0.5. The low correlation means that each attribute by CA has its unique contribution to select summary sentences. The underlying reason for the low correlation may be that, the reasons written by the volunteers help the requesters easily understand the meaning of each candidate attribute. Based on these reasons, requesters merge the candidate attributes in advance under the guidance of Heuristic Construction Rules in Fig. 5.

**Conclusion**: CA can construct meaningful attributes with domain specific knowledge. The attributes constructed by CA for bug report summarization have low correlations.

*2) Answer to RQ2:* In this part, the 11 new attributes by CA are fed into LRCA. We investigate whether LRCA could improve the existing approaches for bug report summarization.

In Fig. 10, we summarize the experimental results of all algorithms with a bar chart in terms of four metrics, including *Precision*, *Recall*, *F-score*, and *Pyramid*. Below the bar chart, we also present the exact values of these algorithms. As shown in Fig. 10, LRCA consistently outperforms all the comparative algorithms in terms of every metric under the paired Wilcoxon signed rank test (p<0.05). LRCA improves BRC [15] by 1.33%, 10.11%, 8.94%, 5.89% in terms of *Precision*, *Recall*, *F-score*, and *Pyramid*, respectively. Since LRCA and BRC employ the same model (logistic regression) with distinct sets of attributes, it implies that the attributes by CA in LRCA contribute to the success of LRCA.

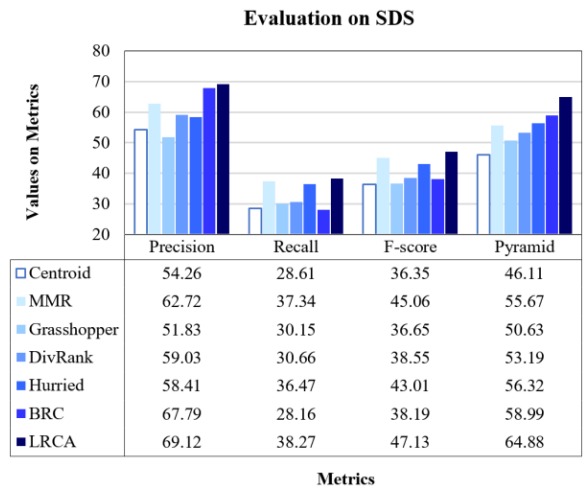

Fig. 10.  Experimental results of algorithms over SDS.

| | Precision | Recall | F-score | Pyramid |
|---|---|---|---|---|
| Centroid | 54.26 | 28.61 | 36.35 | 46.11 |
| MMR | 62.72 | 37.34 | 45.06 | 55.67 |
| Grasshopper | 51.83 | 30.15 | 36.65 | 50.63 |
| DivRank | 59.03 | 30.66 | 38.55 | 53.19 |
| Hurried | 58.41 | 36.47 | 43.01 | 56.32 |
| BRC | 67.79 | 28.16 | 38.19 | 58.99 |
| LRCA | 69.12 | 38.27 | 47.13 | 64.88 |

When comparing LRCA against unsupervised algorithms, LRCA improves Hurried [17] by 10.71% in terms of *Precision*. Meanwhile, LRCA also outperforms Hurried by 1.8% - 8.56% in the other three metrics. Out of the four unsupervised algorithms proposed in [18], namely Centroid, MMR, Grasshopper, and DivRank, MMR performs best in all the four metrics. However, LRCA could also outperform MMR in every metric. For example, the value of *Pyramid* achieved by LRCA is 64.88% whereas that of MMR is 55.67%. For the other three metrics, LRCA also performs better than MMR.

Based on the above observations, we can conclude that LRCA with new attributes by CA can consistently outperform the comparative algorithms. It indicates that CA could provide effective attributes for bug report summarization.

**Conclusion**: LRCA performs better than all the existing approaches. CA is a good way to construct effective attributes for bug report summarization.

*3) Answer to RQ3:* In this part, we investigate the impact of the number of volunteers on the effectiveness of LRCA.

We first partition the volunteers into distinct sized combinations and achieve each combination's contributed attributes. The combination's contributed attributes are referred to as the attributes, which are constructed under the inspiration of the reasons submitted by the volunteers in this combination. Then, we evaluate the performance of LRCA with these attributes to examine the impact of the number of volunteers.

In this study, we employ 21 volunteers to summarize bug reports and provide the reasons for their decisions. Considering that two volunteers provide no reason, we investigate the impact of the combinations of the remaining 19 volunteers.

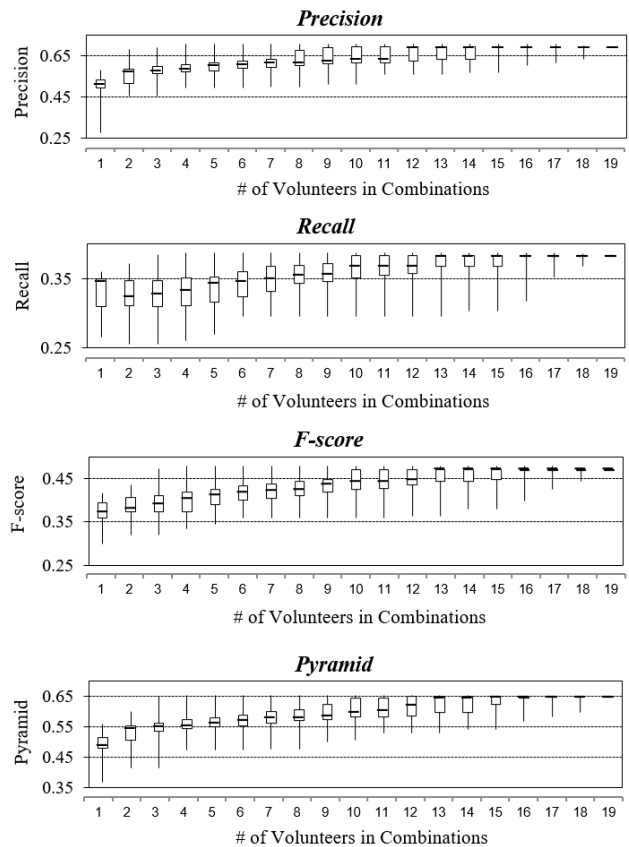

Fig. 11.  Results of LRCA with distinct sets of attributes on SDS.

We partition all the 19 volunteers into the combinations consisting of 1, 2, 3, . . . , 19 volunteers. For example, we have $\binom{19}{3} = 19 \times 18 \times 17/3! = 969$ combinations with 3 volunteers and $\binom{19}{4} = 3876$ combinations with 4 volunteers. For each combination, we evaluate the performance of LRCA with its contributed attributes over SDS. Although we have $\sum_{i=1}^{19} \binom{19}{i}$ distinct combinations, it should be noted that many combinations contribute to the same set of attributes. Therefore, we only need to evaluate the performance of LRCA with 113 distinct sets of attributes.

In Fig. 11, we present the results of LRCA with the boxplots over the equally sized combinations. In this figure, the metrics achieved by LRCA are plotted against the number of volunteers within each combination. As shown in Fig. 11, the median values of the metrics gradually increase and retain relatively stable along with the growth of the number of volunteers. For example, the median values of *Recall* keep nearly the same when more than 13 volunteers are involved. We can also observe similar trends in terms of *Precision*, *F-Score*, and *Pyramid*. Meanwhile, the ranges of the boxplots gradually decrease along with the growth of the number of volunteers. It indicates that LRCA could usually achieve better performance when involving more volunteers. However, the impact of the number of volunteers gradually declines along with the growth of the number of volunteers. In this study, we also use significant level to demonstrate the difference of different combinations. We use the median values of the evaluation metrics to denote the performance of different sized combinations. We find that when the number of volunteers increases to 13, we can achieve no significant different performance with all 19 volunteers on all the evaluation metrics under the paired Wilcoxon signed rank test (p<0.05).

TABLE VI
WINNING TABLE OF LRCA AGAINST COMPARATIVE ALGORITHMS OVER *Precision*.

| Win% | | | | | | | | # of Volunteers Combinations | | | | | | | | | | | |
|---|---|---|---|---|---|---|---|---|---|---|---|---|---|---|---|---|---|---|---|
| | 1 | 2 | 3 | 4 | 5 | 6 | 7 | 8 | 9 | 10 | 11 | 12 | 13 | 14 | 15 | 16 | 17 | 18 | 19 |
| Centroid | 21.1 | 63.2 | 83.5 | 92.6 | 96.9 | 98.9 | 99.7 | 99.9 | 100.0 | 100.0 | 100.0 | 100.0 | 100.0 | 100.0 | 100.0 | 100.0 | 100.0 | 100.0 | 100.0 |
| MMR | 0.0 | 1.2 | 3.6 | 7.3 | 12.2 | 18.4 | 25.7 | 34.0 | 43.2 | 52.7 | 62.3 | 71.4 | 79.6 | 86.7 | 92.2 | 96.3 | 98.8 | 100.0 | 100.0 |
| Grasshoper | 36.8 | 71.9 | 86.7 | 94.2 | 97.7 | 99.2 | 99.8 | 100.0 | 100.0 | 100.0 | 100.0 | 100.0 | 100.0 | 100.0 | 100.0 | 100.0 | 100.0 | 100.0 | 100.0 |
| DivRank | 0.0 | 17.5 | 30.2 | 45.5 | 60.1 | 72.1 | 81.3 | 87.9 | 92.6 | 95.7 | 97.7 | 98.9 | 99.6 | 99.9 | 100.0 | 100.0 | 100.0 | 100.0 | 100.0 |
| Hurried | 0.0 | 18.7 | 35.3 | 53.4 | 68.5 | 79.3 | 86.6 | 91.4 | 94.6 | 96.8 | 98.2 | 99.1 | 99.6 | 99.9 | 100.0 | 100.0 | 100.0 | 100.0 | 100.0 |
| BRC | 0.0 | 0.0 | 0.9 | 3.2 | 6.6 | 11.0 | 16.2 | 22.2 | 28.9 | 36.3 | 44.1 | 52.1 | 60.1 | 68.0 | 75.5 | 82.5 | 88.9 | 94.7 | 100.0 |

TABLE VII
WINNING TABLE OF LRCA AGAINST COMPARATIVE ALGORITHMS OVER *Recall*.

| Win% | | | | | | | | # of Volunteers Combinations | | | | | | | | | | | |
|---|---|---|---|---|---|---|---|---|---|---|---|---|---|---|---|---|---|---|---|
| | 1 | 2 | 3 | 4 | 5 | 6 | 7 | 8 | 9 | 10 | 11 | 12 | 13 | 14 | 15 | 16 | 17 | 18 | 19 |
| Centroid | 89.5 | 95.9 | 99.3 | 99.9 | 100.0 | 100.0 | 100.0 | 100.0 | 100.0 | 100.0 | 100.0 | 100.0 | 100.0 | 100.0 | 100.0 | 100.0 | 100.0 | 100.0 | 100.0 |
| MMR | 0.0 | 0.0 | 1.0 | 3.0 | 5.7 | 8.9 | 12.7 | 17.2 | 22.8 | 29.4 | 37.0 | 45.4 | 54.4 | 63.5 | 72.4 | 80.8 | 88.3 | 94.7 | 100.0 |
| Grasshoper | 78.9 | 91.2 | 89.3 | 87.9 | 89.5 | 92.3 | 95.1 | 97.2 | 98.6 | 99.4 | 99.8 | 99.9 | 100.0 | 100.0 | 100.0 | 100.0 | 100.0 | 100.0 | 100.0 |
| DivRank | 78.9 | 83.6 | 81.9 | 81.9 | 84.1 | 87.2 | 90.3 | 93.0 | 95.3 | 97.0 | 98.3 | 99.1 | 99.6 | 99.9 | 100.0 | 100.0 | 100.0 | 100.0 | 100.0 |
| Hurried | 0.0 | 0.6 | 2.9 | 7.0 | 12.5 | 19.1 | 26.5 | 34.4 | 42.8 | 51.4 | 60.1 | 68.5 | 76.5 | 83.7 | 89.9 | 94.8 | 98.2 | 100.0 | 100.0 |
| BRC | 89.5 | 95.9 | 99.3 | 99.9 | 100.0 | 100.0 | 100.0 | 100.0 | 100.0 | 100.0 | 100.0 | 100.0 | 100.0 | 100.0 | 100.0 | 100.0 | 100.0 | 100.0 | 100.0 |

TABLE VIII
WINNING TABLE OF LRCA AGAINST COMPARATIVE ALGORITHMS OVER *F-score*.

| Win% | | | | | | | | # of Volunteers Combinations | | | | | | | | | | | |
|---|---|---|---|---|---|---|---|---|---|---|---|---|---|---|---|---|---|---|---|
| | 1 | 2 | 3 | 4 | 5 | 6 | 7 | 8 | 9 | 10 | 11 | 12 | 13 | 14 | 15 | 16 | 17 | 18 | 19 |
| Centroid | 68.4 | 89.5 | 91.0 | 91.5 | 93.3 | 95.5 | 97.3 | 98.5 | 99.3 | 99.7 | 99.9 | 100.0 | 100.0 | 100.0 | 100.0 | 100.0 | 100.0 | 100.0 | 100.0 |
| MMR | 0.0 | 0.0 | 0.4 | 1.5 | 3.6 | 6.8 | 11.0 | 16.4 | 22.8 | 30.0 | 38.0 | 46.5 | 55.3 | 64.2 | 72.8 | 80.9 | 88.3 | 94.7 | 100.0 |
| Grasshoper | 63.2 | 83.0 | 87.4 | 90.1 | 92.9 | 95.4 | 97.2 | 98.5 | 99.3 | 99.7 | 99.9 | 100.0 | 100.0 | 100.0 | 100.0 | 100.0 | 100.0 | 100.0 | 100.0 |
| DivRank | 36.8 | 44.4 | 55.9 | 67.9 | 77.6 | 84.5 | 89.3 | 92.8 | 95.2 | 97.0 | 98.3 | 99.1 | 99.6 | 99.9 | 100.0 | 100.0 | 100.0 | 100.0 | 100.0 |
| Hurried | 0.0 | 1.2 | 5.6 | 12.0 | 19.8 | 28.5 | 37.8 | 47.3 | 56.7 | 65.7 | 73.9 | 81.2 | 87.3 | 92.1 | 95.7 | 98.0 | 99.4 | 100.0 | 100.0 |
| BRC | 42.1 | 46.8 | 57.2 | 68.5 | 77.9 | 84.7 | 89.4 | 92.8 | 95.2 | 97.0 | 98.3 | 99.1 | 99.6 | 99.9 | 100.0 | 100.0 | 100.0 | 100.0 | 100.0 |

TABLE IX
WINNING TABLE OF LRCA AGAINST COMPARATIVE ALGORITHMS OVER *Pyramid*.

| Win% | | | | | | | | # of Volunteers Combinations | | | | | | | | | | | |
|---|---|---|---|---|---|---|---|---|---|---|---|---|---|---|---|---|---|---|---|
| | 1 | 2 | 3 | 4 | 5 | 6 | 7 | 8 | 9 | 10 | 11 | 12 | 13 | 14 | 15 | 16 | 17 | 18 | 19 |
| Centroid | 84.2 | 98.2 | 99.9 | 100.0 | 100.0 | 100.0 | 100.0 | 100.0 | 100.0 | 100.0 | 100.0 | 100.0 | 100.0 | 100.0 | 100.0 | 100.0 | 100.0 | 100.0 | 100.0 |
| MMR | 5.3 | 18.1 | 32.8 | 47.5 | 59.9 | 69.8 | 77.9 | 84.3 | 89.5 | 93.4 | 96.2 | 98.1 | 99.2 | 99.7 | 99.9 | 100.0 | 100.0 | 100.0 | 100.0 |
| Grasshoper | 36.8 | 76.0 | 89.4 | 95.0 | 97.8 | 99.1 | 99.7 | 99.9 | 100.0 | 100.0 | 100.0 | 100.0 | 100.0 | 100.0 | 100.0 | 100.0 | 100.0 | 100.0 | 100.0 |
| DivRank | 21.1 | 60.8 | 80.9 | 89.1 | 93.0 | 95.5 | 97.3 | 98.5 | 99.3 | 99.7 | 99.9 | 100.0 | 100.0 | 100.0 | 100.0 | 100.0 | 100.0 | 100.0 | 100.0 |
| Hurried | 0.0 | 11.1 | 22.0 | 35.6 | 49.3 | 61.7 | 72.2 | 80.8 | 87.4 | 92.4 | 95.8 | 97.9 | 99.1 | 99.7 | 99.9 | 100.0 | 100.0 | 100.0 | 100.0 |
| BRC | 0.0 | 1.2 | 4.2 | 9.4 | 16.1 | 23.8 | 32.2 | 40.9 | 49.7 | 58.3 | 66.7 | 74.5 | 81.5 | 87.7 | 92.6 | 96.4 | 98.8 | 100.0 | 100.0 |

We also present the winning tables of LRCA against each existing approach in Table VI - IX. Each cell in the winning tables for $k$ volunteers indicates the portion of combinations with $k$ volunteers providing effective attributes such that LRCA outperforms its comparative algorithms. In these tables, we mark all the cells with values larger than 50% in dark. For example, 52.1% of the combinations with 12 volunteers could contribute the attributes with which LRCA outperforms BRC in terms of *Precision*. When we consider the metric of *Precision*, over 50% combinations with more than 12 volunteers could contribute effective attributes with which LRCA outperforms all the comparative approaches. For all the four metrics, we can conclude that when more than 13 volunteers are involved, over 50% combinations could have LRCA achieve better performance than all the baselines.

In the following part, we investigate the underlying reasons of the behaviors of LRCA for distinct sized combinations. In Table. X, we list the attributes contributed by the volunteers. For each volunteer, we use one column of entries to present her/his contributed attributes. As shown in Table. X, we have the following observations.

1) No one could contribute to all the attributes. The vol- unteers' capabilities of contributing attributes vary from volunteer to volunteer.

2) It is easier to achieve some attributes than other at- tributes. In Table X, only two volunteers (V6 and V7) contribute to the attribute CODE while 13 volunteers contribute to the attribute SWT.

#### TABLE X
DISTRIBUTION OF ATTRIBUTES BY VOLUNTEERS.

| | | | | | | | | | | | Volunteer | | | | | | | | | |
|---|---|---|---|---|---|---|---|---|---|---|---|---|---|---|---|---|---|---|---|---|
| | | V1 | V2 | V3 | V4 | V5 | V6 | V7 | V8 | V9 | V10 | V11 | V12 | V13 | V14 | V15 | V16 | V17 | V18 | V19 |
| Attribute | SWT | ● | ● | ● | | | ● | ● | | ● | ● | ● | | | ● | | | | ● | ● |
| | SWD | | | | | | ● | | | | | | | | | | | | ● | |
| | DUP | | ● | | ● | | | | | | | | | | | | | | | |
| | SLEN | | | | | ● | | ● | ● | | | | ● | | ● | ● | ● | | | |
| | SI | ● | ● | ● | | ● | | | ● | | ● | ● | ● | | | | ● | ● | ● | |
| | SLOC | ● | ● | ● | | ● | ● | | ● | | ● | ● | ● | ● | ● | ● | ● | | | |
| | CLEN | | | | | | | | | | | | | | | ● | | | | |
| | DES | | | | | | | | | | | | | | | | | | | |
| | CCW | ● | | | | | ● | ● | | ● | | ● | ● | ● | | | | ● | ● | ● |
| | CODE | | | | | | | ● | ● | ● | | | | | | | | | | |
| | REP | | | | | | | | | | ● | | | | | | | | | |

#### TABLE XI
STATISTICS OF BUG REPORTS IN BRSBS.

| Data Set | Project | # of Bug Reports | Max. # of Sentences | Min # of Sentences | Aver. # of Sentences |
|---|---|---|---|---|---|
| BRSB1 | Mozilla | 49,319 | 4,993 | 1 | 45.23 |
| BRSB2 | Eclipse | 25,696 | 3,184 | 1 | 35.54 |
| BRSB3 | KDE | 21,039 | 5,287 | 1 | 51.10 |
| BRSB4 | Gnome | 9,123 | 4,409 | 1 | 51.39 |

3) Volunteers are able to compensate with each other to inspire the requesters to construct more attributes, until all the attributes are gradually covered.

Based on the above observations, when involving more volunteers, more discriminative attributes can be achieved by CA. Hence, the results of LRCA gradually improve and retain stable along with the growth of the number of volunteers.

**Conclusion**: LRCA with new attributes could achieve better performance when involving more volunteers. Along with the growth of the number of volunteers, its impact on LRCA gradually declines.

*4) Answer to RQ4:* In this subsection, we compare LRCA against the existing algorithms over large data sets.

Since it is time-consuming and tedious to annotate a large number of bug reports, the publicly available data set SDS only contains 36 annotated bug reports. To evaluate the performance of algorithms over large data sets, we build a series of new large data sets Bug Report Summarization Benchmarks (BRSBs) with 105,177 bug reports in an alternative.

As to the literature [17], [20], the title of a bug report can be a good high-level summary of this bug report. Developers usually first focus on the bug report title to understand the topic of the bug report, and then read the important sentences (typically, the selected sentences by an algorithm) to find the main problems and solutions of the bug report [17]. Inspired by this phenomenon, given a bug report, we could inject its title into the contents of this bug report, namely the concatenation of the bug report's description and comments, to form a revised bug report. Then we can evaluate each algorithm by checking whether the algorithm can detect the title in the resulting summary from the revised bug report. If the title is detected, it means that the algorithm has the ability to extract such high-level summary sentences from a bug report.

We download bug reports from the same four open source projects as in SDS, namely Mozilla, Eclipse, KDE, and Gnome. Table XI presents the statistical information about these downloaded bug reports. We use the bug reports submitted to Mozilla, Eclipse, KDE in 2008 which are marked as 'fixed' to create the BRSBs, since nearly 40% of the bug reports in SDS are submitted in 2008. However, the Gnome bug repository is closed after 2005, so we choose the latest fixed bug reports, namely the reports submitted in 2005 to create the BRSDs. Follow this criterion, we collect 25,696 reports for Eclipse, 49,319 reports for Mozilla, 21,039 reports for KDE and 9,123 reports for Gnome as the new data sets. In these data sets, the title of every bug report is injected to a random location among the description/comments as a specific "comment" (provided by the reporter) so as to form a revised bug report. The underlying reason for this action is that we find that a bug report usually presents conversations in a relatively loose form. For example, a user may inject a comment in the conversation to assess or conclude the sentences several comments before [9]. Thus, the injected title can be viewed as an assessment or conclusion of previous comments said by the reporter.

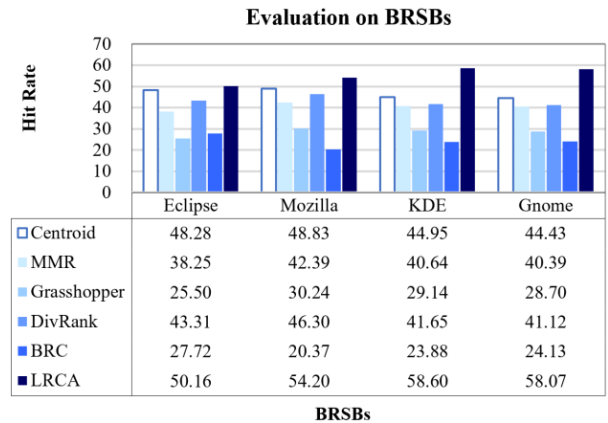

Fig. 12. Experimental results of algorithms over BRSBs.

Since there exists no manually annotated gold standard summary, we evaluate algorithms in terms of *HitRate*.

$$HitRate = Num_{hit}/Num_{report}, \quad (6)$$

where $Num_{hit}$ is the number of bug reports whose titles are successfully retrieved by algorithms and $Num_{report}$ is the total number of bug reports in every new data set.

For the comparative algorithms, Hurried employs the title of a bug report to calculate one of its attributes. Hence, for a fair comparison, we only run the other comparative algorithms on BRSBs. Fig. 12 summarizes the results.

As shown in Fig. 12, Centroid performs best among the four unsupervised algorithms. The values of *HitRate* achieved by Centroid are 48.28%, 48.83%, 44.95%, and 44.43% over Mozilla, Eclipse, KDE, and Gnome data sets, respectively. In contrast, the values of *HitRate* achieved by Grasshopper are 25.50%, 30.24%, 29.14%, and 28.70%. When compared against the best unsupervised approach, LRCA improves

---

[9] In Eclipse bug report 214067, a user adds comment 6 to assess a far away comment 2 (https://bugs.eclipse.org/bugs/show_bug.cgi?id=214067).

Q5: *Do you like to complete the task with the reward of a USB flash disk?*

• 5: like it very much.

• 4: like it.

• 3: not sure.

• 2: dislike it.

• 1: hate the task.

Fig. 13. Question related to the interests of volunteers.

TABLE XII
SURVEY RESULTS.

| Content | Result | Notes |
|---|---|---|
| Q3: Readability of Bug Reports | 3.75 | Point 5 means that it is very easy to understand |
| Q5: Interest in the Task | 4.45 | Point 5 means that one likes the task very much |

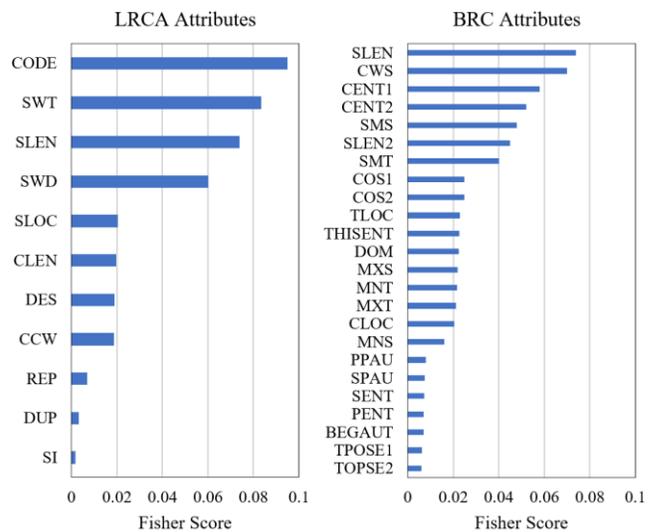

Fig. 14. The discriminability of attributes in LRCA and BRC.

Centroid by 1.88%-13.65% over distinct data sets. For the other unsupervised approaches, LRCA improves them by up to 29.46% in terms of *HitRate*. When compared against the existing supervised approach BRC, LRCA significantly improves BRC by 22.44%-34.72% over these new data sets. The underlying reason of the poor performance of BRC may be that each bug report in BRSBs only has one summary sentence (the injected title) and many negative instances (all the other sentences). The imbalanced data sets may affect the performance of the attributes in training BRC. In contrast, the attributes in LRCA can handle both the manually annotated data set SDS and the automatically created data sets BRSBs.

**Conclusion**: LRCA outperforms the competitive algorithms on the large data sets BRSBs. The attributes constructed by CA can handle different types of data sets.

*5) Answer to RQ5:* In this part, we analyze the feasibility of recruiting interested volunteers with the necessary knowledge for CA.

In this study, we invite 450 college students to participate in CA. We successfully recruit 21 volunteers, i.e., around 5% of college students are involved. It indicates that directly recruiting students for crowdsourcing SE tasks is a practical way. Prior study in the literature also suggests that college students inhabit a large amount of volunteers in a crowdsourcing scenario [52]. Furthermore, we ask the volunteers in this study about their interests in completing the tasks (see Q5 in Fig. 13). The results are presented in Table XII. The average interest of the volunteers to this task is 4.45, i.e., these volunteers enjoy such tasks with a reward. The previous study [53] also proves that a reward can attract volunteers to participate in crowdsourcing tasks. Hence, CA can recruit interested volunteers to complete an SE task.

In addition, CA requires little domain knowledge of the volunteers. In this study, we ask the volunteers to evaluate the readability of crowdsourcing bug reports (Q3 in Fig. 8). If a volunteer marks point 5 on this question, it means that the bug reports are very easy for him/her to understand. Although these volunteers have never conducted researches on bug reports, most volunteers can easily understand the bug reports with the average readability score of 3.75 (in Table XII). The result means that it is possible to recruit volunteers with the necessary knowledge to understand CA tasks by inviting college students.

**Conclusion**: We can employ interested volunteers with sufficient knowledge to participate in CA.

*F. Summary of RQs*

Inspired by a group of volunteers, LRCA with crowdsourced attributes could well summarize bug reports against the comparative approaches. The success of LRCA also validates that crowdsourcing is helpful to facilitate resolving SE tasks.

## VII. DISCUSSION

*A. Attribute Comparison*

Since both BRC and LRCA propose a set of attributes for bug report summarization, we compare these attributes in Fig. 14. In the figure, we rank the attributes from different algorithms according to the attribute discriminability. The discriminability is computed by *Fisher-score* [54], a score to evaluate the attributes discriminability in supervised algorithms [15]. Given a set of instances $x_1, x_2, \ldots, x_m$ with $n_+$ positive and $n_-$ negative instances, the $i$th attribute's value of *Fisher-score* is computed as:

$$Fisher\text{-}score(i) = \frac{(\bar{x}_i^+ - \bar{x}_i)^2 + (\bar{x}_i^- - \bar{x}_i)^2}{\frac{1}{n_+-1}\sum_{k=1}^{n_+}(x_{k,i}^+ - \bar{x}_i^+)^2 + \frac{1}{n_--1}\sum_{k=1}^{n_-}(x_{k,i}^- - \bar{x}_i^-)^2}, \quad (7)$$

where $\bar{x}_i$, $\bar{x}_i^+$, $\bar{x}_i^-$ are the average values of the whole, the positive, and the negative instances' $i$th attributes, respectively. $x_{k,i}^+$ is the value of the $k$th positive instance's $i$th attribute, and $x_{k,i}^-$ is the value of the $k$th negative instance's $i$th attribute. According to the formula, the attributes with higher values of *Fisher-score* are more capable of discriminative instances. We calculate *Fisher-score* based on the human annotated data set SDS and sort all the attributes in descending order in Fig. 14.

As shown in Fig. 14, both LRCA and BRC utilize many discriminative attributes to summarize bug reports. CODE, SWT, and SLEN are the top three most helpful attributes in LRCA, while SLEN, CWS, and CENT1 are important to BRC. Some attributes constructed by CA are also utilized in BRC, including SLEN, SLOC (named as CLOC in BRC), and REP (named as BEGAUT in BRC). These attributes have different discriminability. For example, SLEN is much more discriminative than REP (or BEGAUT) in both attribute sets. These overlapped attributes are generic conversation-based attributes to measure the sentence length, sentence location and the reporter of a bug report. Although BRC proposes 24 generic conversation-based attributes, CA also identifies other helpful generic attributes for bug report summarization, e.g., SWT, DUP, and SI. In addition, after manually summarizing bug reports by the volunteers, domain-specific attributes can also be identified from the volunteers' reasons, e.g., CODE and SWD. These domain-specific attributes have high discriminability to detect summary sentences.

Another difference between attributes in BRC and LRCA is that, many attributes in BRC have similar meaning or discriminability. For example, BRC measures *Tprob* with similar attributes MNT and MXT. In contrast, the correlation between LRCA attributes is low as shown in Fig. 9. Most correlation values are below 0.5. The attributes SWD and DES are partially correlated with a correlation value of 0.72. However, when we remove one of the two attributes, e.g., DES, *F-score* slightly increases from 0.4713 to 0.4779 and *Pyramid* drops from 0.6488 to 0.6436, which means DES is still useful to some evaluation metrics.

To conclude, CA could identify unique attributes that are different from previous studies. Some unique attributes have high discriminability to detect summary sentences.

### B. Requester Background

This section investigates the influence of requesters' background knowledge on attributes construction. We first compare the constructed attributes by the two requesters, and then show the influence of requesters on bug report summarization.

In this study, two requesters construct attributes from volunteers' reasons independently. Req1 has a research experience in software engineering for 3 years and Req2 has researched on this area for 7 years. The two requesters achieve a concordance rate of 76.5% on attribute construction as mentioned in Section V-A. The conflicts are mainly from two parts:

1) the conflicts on selecting candidate attributes. Requesters may select different terms or phrases as candidate attributes from the reasons. For example, for the reason "this sentence explains a best way of solving the problem", the object of the reason is "the best way of solving the problem". Req1 constructs attributes from the term "problem". He utilizes 0 and 1 to represent whether a sentence contains this term (C-CW), since the term "problem" also exists in the corresponding sentence of the bug report. In contrast, Req2 constructs the attribute SWD from this reason as explained in Section V-B. However, we find that the influence on such conflicts may be minimized in the crowdsourcing setting. Req1 could also construct the attribute SWD from the reason "it is similar with the sentences in the bug report" (in Section V-B).

TABLE XIII
RESULTS ON LRCA, $LRCA_{Req1}$ AND $LRCA_{Req2}$.

| Algorithm | Precision | Recall | F-score | Pyramid |
|---|---|---|---|---|
| $LRCA_{Req1}$ | 68.67 | 36.52 | 45.53 | 62.43 |
| $LRCA_{Req2}$ | 70.32 | 36.76 | 46.20 | 64.04 |
| LRCA | 69.12 | 38.27 | 47.13 | 64.88 |

2) the conflicts on calculating the attributes. Requesters may calculate candidate attributes with different metrics. We explain all the differences in attribute calculation as follows:

[SWT] Req1 utilizes words in the entire bug reports to represent the bug report topic, while Req2 regards the top 20% words with highest TF-IDF values as the bug report topic.

[SWD] Req1 directly measures the similarity between the current sentence and the description of the bug report, while Req2 sets the sentences in bug report description to 1.

[CODE] Two requesters detect code snippets with different heuristic patterns. They merge the heuristic patterns to achieve the final ones.

For the above two conflicts, the requesters' background knowledge has small influence on the candidate attribute selection, since all the attributes could be detected when the number of reasons is large. However, the background knowledge may influence the way to calculate attributes. The following part discusses this influence on bug report summarization.

We construct two additional algorithms named $LRCA_{Req1}$ and $LRCA_{Req2}$, which utilizes the attributes constructed from Req1 and Req2 respectively. Table XIII shows the results on applying LRCA, $LRCA_{Req1}$ and $LRCA_{Req2}$ on the SDS data set. As shown in Table XIII, when only utilizing the attributes from Req1, all the evaluation metrics drop, e.g., *Precision* drops from 69.12 to 68.67. The attributes from Req2 are slightly better, which outperforms $LRCA_{Req1}$ from 0.24% to 1.65% on different evaluation metrics. The reason may be that Req2 has 4 years more research experience on software engineering. After pair-wise discussion, most of the evaluation metrics increase with the final attributes.

To conclude, the requesters' background knowledge has small influence on the candidate attribute selection. It mainly influences the attribute calculation, resulting in different performances on bug report summarization.

### C. Attributes Construction by Related Work

This section investigates whether we can achieve effective attributes by reviewing related work for bug report summarization. To answer this question, we identify the attributes in related work and propose an algorithm named *Combine* for comparison which combines the attributes from related work.

As discussed in Section VI-C, besides the 24 attributes in BRC, the unsupervised algorithms also have attribute-related elements. The algorithm *Hurried* selects a summary sentence with consideration of its similarity with the bug report title, the sentences in the bug report description, and the sentiment of a sentence. We construct three attributes from this algorithm.

TABLE XIV
RESULTS ON LRCA AND COMBINE.

| Algorithm | Precision | Recall | F-score | Pyramid |
|-----------|-----------|--------|---------|---------|
| Combine   | 64.31     | 30.02  | 39.72   | 59.84   |
| LRCA      | 69.12     | 38.27  | 47.13   | 64.88   |

· We calculate the similarity between the current sentence and the bug report title with VSM.

· We use 0 and 1 to represent whether a sentence is in the bug report description.

· We detect the sentiment of a sentence [55], and use 1, 0, -1 to denote the positive, neutral, negative sentiment.

In addition, the algorithms *Centroid*, *MMR*, *Grasshopper*, and *DivRank* use words and sentences as attributes. Since there are more than 2,000 sentences in the SDS data set, the high-dimension and sparse attributes are usually ineffective for classification [35]. We remove these attributes. At last, there are 27 attributes for *Combine*. We feed these attributes into the same framework as LRCA for a fair comparison.

As shown in Table XIV, LRCA significantly outperforms the algorithm *Combine*. LRCA improves *Combine* by 4.81%, 8.25%, 7.41% and 5.04% with respective to *Precision*, *Recall*, *F-score* and *Pyramid* respectively. The reason is that some attributes in previous studies may have negative influence on *Combine*. Besides, CA also identifies several discriminative attributes to improve the performance of LRCA.

To conclude, since researchers tend to combine several methods together to construct attributes as to the survey in Section III, CA is promising to construct more effective attributes after researchers utilizing the traditional methods, e.g., knowledge transfer, heuristic or experience, etc.

### D. Student Volunteers

We discuss whether student volunteers can be regarded as representatives to draw conclusions for CA. We recruit students to draw conclusions for three reasons.

First, we find that students can represent the volunteers in crowdsourcing scenarios, since students inhabit a large number of volunteers in crowdsourcing platforms [52].

Second, student groups are widely used in previous studies for crowdsourcing [56], [57]. Researchers also show that, for some SE tasks, students and professionals may achieve similar performance except for minor differences [58]. Since CA asks volunteers to fill in reasons for a summary sentence instead of deciding the final attributes, students may have acceptable performance.

Third, in this study, we also consider the background knowledge of students. For a crowdsourcing task related to SE, we require that all the students have the background knowledge of computer science and English.

Hence, we use student groups as volunteers to draw conclusions for the attribute construction methods CA.

## VIII. THREATS TO VALIDITY

### A. Generality

The generality of CA should be further studied. In this study, we evaluate CA with more than 100,000 bug reports from four desktop software. With the increasing number of mobile Apps [30], we also plan to apply CA to summarize bug reports of mobile Apps in the future. Besides bug report summarization, CA can also be leveraged to facilitate many other SE tasks over textual SE data, such as severity prediction [59] and duplicate bug reports detection [8]. For example, in the SE task of severity prediction, requesters can employ some volunteers to manually check the severity of a population of bug reports and ask them to provide the reasons in making their decisions. With such reasons, requesters could construct new attributes. Hence, we plan to extend CSEP to support more types of SE data. With the extended CSEP, we can evaluate CA with more SE tasks thoroughly against all the existing attribute construction methods.

### B. Subjectiveness

Attribute construction is a subjective process and the effectiveness of CA might be influenced by the requesters, especially for deriving adequate measures for attributes. Since two requesters in this study achieve the concordance rate of 76.5% in constructing attributes for bug report summarization, it implies that CA with properly defined construction rules could work well in achieving new attributes for the SE task. In addition, requesters employ several naive methods to measure attributes, including VSM, boolean values, etc., to alleviate the influence of sophisticated measurement methods.

Besides, the background knowledge of requesters is a threat to the effectiveness of CA. In this study, we employ two requesters to construct effective attributes from the reasons by volunteers. Since both requesters have a research experience on SE for more than three years, there is a threat that poor quality attributes may be constructed when requesters have little background knowledge, e.g., they have no research experience on SE. Since experienced researchers still prefer combining several methods to construct attributes according to our survey, CA is helpful for them.

Another threat is the bias on evaluating the requesters' background in Section VII-B. We ask the authors of this study to act as the requesters, which may bring a bias on attribute construction. To alleviate this threat, we do not predefine coding schemes or possible attributes for bug report summarization. The requesters independently infer attributes from reasons under HCRs. Hence, it demonstrates that with a set of construction rules, crowd-generated data in crowdsourcing are crucial resources to infer attributes for SE tasks.

To better evaluate these threats, in future work, we plan to recruit more requesters of different experience and compare the differences of them in constructing attributes for the same groups of reasons. In addition, we also plan to automate CA for large numbers of reasons. We find that many parts of HCRs can be automated, including analyzing the part-of-speech to identify candidate attributes, grouping candidate attributes by synonyms, etc. These parts can be automated by some natural language processing techniques.

### C. Crowdsourcing process

As a type of crowdsourcing, CA may be impacted by some other factors, including the quality of answers and the design of questions, etc. These factors have been investigated in the research of traditional crowdsourcing. Requesters can better control the quality of answers with gold standard data when volunteers process the task [60]. In addition, a well-designed question can be achieved by following some guidelines [61]. These methods for traditional crowdsourcing could be employed to further improve CA.

Meanwhile, the quantity and quality of the volunteers are the threats to crowdsourcing process. For the quantity, by inviting college students, CA can attract tens of volunteers to participate in. In addition, the results of supervised algorithms with new attributes constructed by CA gradually become stable along with the growth of the number of volunteers. For the quality, solid SE knowledge is not mandatory for every individual volunteer, since volunteers could complement with each other in the process of CA.

### IX. Related Work

In this section, we review the related work of this study, including attribute engineering and crowdsourcing in SE.

### A. Attribute Engineering in MSR

Attribute engineering means feature engineering in the societies of data mining and machine learning, which includes attribute construction, attribute selection and attribute extraction [31]–[33]. Although no systematic work has been conducted on attribute construction in MSR, a few studies have been performed on both attribute selection and attribute extraction for facilitating SE tasks.

Yang et al. show that three attribute selection schemes (Information Gain, Chi-Square, and Correlation Coefficient) can improve severity prediction of defect reports on test cases from Eclipse and Mozilla [37]. Shivaji et al. investigate several attribute selection schemes to substantially reduce the number of attributes and achieve significant improvement on the performance of Naïve Bayes and support vector machine over the task of code change-based bug prediction [34]. Xuan et al. combine instance selection and attribute selection to reduce software data and show that data reduction can effectively improve the accuracy of bug triage [35].

In contrast, Turhan et al. employ two attribute extraction techniques, namely principal component analysis and Isomap, for extracting new attributes from existing ones and evaluate these methods with support vector regression on the SE task of software cost estimation [36].

The above studies are based on the initial set of attributes. However, none of the above studies discusses the methods for constructing such initial attributes. In this study, we investigate how to employ crowdsourcing to construct new attributes for a typical SE task, namely bug report summarization.

### B. Crowdsourcing in Software Engineering

Recent years have witnessed the growing research of crowdsourcing in SE. In general, a crowdsourcing method breaks an SE task into some subtasks and assigns them to crowd for solving. Kazman and Chen present an overview analysis on how crowdsourcing changes the future of SE [29]. Recently, Mao et al. survey the crowdsourcing usage in SE [62].

A lot of research work has been conducted to bring crowdsourcing into SE processes, including requirements analysis, software design, software development, testing, and maintenance. In requirements analysis, Lim et al. develop a tool called StakeSource to identify stakeholders by mutual recommendation [40]. The proposed model is also able to conduct requirements elicitation and prioritization by asking stakeholders to evaluate requirements proposed by other stakeholders [63]. In software design, LaToza et al. explore the recombination strategies in the process of software design competitions [64]. In the process of software development, Nag et al. divide the project NASA SPHERES into small modules, and outsource them to the crowd in TopCoder [65]. Lin et al. utilize crowdsourcing to capture the expectations of users of whether a sensitive resource can be used for Apps in a given category [66]. In software testing, Chen and Kim successfully leverage crowdsourcing to assist test input generation [41]. Micallef, et al. recruit crowd to assist in testing apps in different mobile hardware scenarios [67]. Besides, the test report is also an important resource for crowdsourcing studies. Feng et al. propose test report prioritization strategies to assist reading crowdsourced test reports [56], [57]. Wang et al. classify test reports to assist crowdsourced testing [68]. In addition, some researchers leverage crowd documentation to facilitate API documentation across software processes [69], [70].

Although much research work has been performed, no related work has been conducted for attribute construction with crowdsourcing. In this study, we attempt to adopt crowdsourcing to construct new attributes in MSR.

### X. Conclusion

Bug report summarization is an essential task in software maintenance. It saves developers/testers' time in understanding the software bugs and their solutions [15], and eventually improves the software quality. In this study, we investigate how to employ crowdsourcing to facilitate attribute construction to improve the task of bug report summarization. We first reveal the existing methods for attribute construction in MSR by a survey. Then, we propose the new method CA to infer effective attributes from the crowd-generated data in crowdsourcing. As to the best of knowledge, it is the first attempt towards constructing new attributes by crowdsourcing data. With CA, we successfully construct 11 attributes and propose the new algorithm LRCA to summarize bug reports. To evaluate the performance of LRCA, we build a series of large data sets BRSBs with 105,177 bug reports. Extensive experiments over both the existing data set SDS with 36 annotated bug reports and BRSBs validate the effectiveness of LRCA.

## Acknowledgment

We greatly thank all the volunteers and the authors of attribute-related papers to participate in this study. We sincerely present this work in memory of Prof.Qinbao Song who offered his life to MSR.